# The implications of Labour's plan to scrap Key Stage 2 tests for Progress 8 and secondary school accountability in England

George Leckie, Lucy Prior and Harvey Goldstein

Centre for Multilevel Modelling and School of Education, University of Bristol, UK

**Address for correspondence**

Professor George Leckie

Centre for Multilevel Modelling

School of Education

University of Bristol

35 Berkeley Square

Bristol

BS8 1JA

United Kingdom

g.leckie@bristol.ac.uk

**Funding details**

This research was supported by the UK Economic and Social Research Council under Grant ES/R010285/1.

# The implications of Labour's plan to scrap Key Stage 2 tests for Progress 8 and secondary school accountability in England

**Abstract**

In England, Progress 8 is the Conservative government's headline secondary school performance and accountability measure. Progress 8 attempts to measure the average academic progress pupils make in each school between their KS2 tests and their GCSE Attainment 8 examinations. The Labour opposition recently announced they would scrap the KS2 tests were they to be elected. Such a move, however, would preclude the publication of Progress 8 and would leave schools to be compared in terms of their average Attainment 8 scores or, at best, their Attainment 8 scores only adjusted for school differences in pupil demographic and socioeconomic characteristics. In this paper, we argue and illustrate empirically that this best-case scenario of an 'Adjusted Attainment 8' measure would prove less fair and meaningful than Progress 8 and therefore a backwards step, especially when Progress 8 itself has been criticised as biased against schools teaching educationally disadvantaged intakes.

**Introduction**

In 2016, the Conservative government introduced Progress 8 as their new headline school performance and accountability measure for all state-maintained secondary schools in England (DfE, 2019a). Progress 8 attempts to measure the average academic progress or improvement pupils make in each school between their end of primary school Key Stage 2 (KS2) tests and their end of secondary school GCSE Attainment 8 examinations. The government argue Progress 8 is the fairest and most meaningful way to compare schools as it implicitly accounts for school differences in pupil KS2 prior attainment when pupils start their secondary schooling (DfE, 2019a). Schools with low scores come under increased scrutiny and intervention from Ofsted, the national school inspectorate, attention which can ultimately lead them to be taken over by other more successful schools or multi-academy trusts. Schools are also held publicly accountable for their Progress 8 scores via national school performance tables (compare-school-performance.service.gov.uk) and via the media's republication of these data (e.g., Telegraph, 2019).

In April 2019, Jeremy Corbyn, Leader of the Labour opposition, outlined his party's forward-looking education policy at the National Education Union (NEU) conference in Liverpool (Labour, 2019a). Crucially, for Progress 8, Corbyn declared (Labour, 2019a):

> So today I can give you this commitment: the next Labour government will scrap primary school SATs for seven and eleven year olds.

SATs is the colloquial term by which the national Key Stage 1 (KS1) and KS2 tests are referred. Scrapping SATs is a move supported by the NEU and the vast majority of its primary school teacher members (Guardian, 2019b). Labour's announcement to scrap the KS2 tests reflects widely held concerns regarding the perverse incentives and negative consequences induced by the KS2 tests as a result of their high-stakes role in the primary and secondary school accountability systems (Foley and Goldstein, 2012; NAHT, 2018; morethanascore.org.uk). Thus, for example, Corbyn argues how the current system places "extreme pressure" on children and "forces teachers to 'teach to the test'" and links this to the current "crisis of teacher retention and recruitment" (Labour, 2019a). Such concerns are not unique to England having also been voiced in the US, Australia and other countries where school performance data directly underpins school accountability systems (Amrein-Beardsley, 2014; NFER, 2018; OECD, 2008; Koretz, 2017). Corbyn goes on to call for "a new system that separates the assessment of schools from the assessment of children" (Labour, 2019a). What Corbyn does not do, however, is to spell out what such a system might look like, simply stating that "We will consult with teaching unions, parents and experts and bring forward proposals for a new system" (Labour, 2019a). Likewise, the NEU also call for a "sensible alternative" to the high-stakes attainment tests, but give few further details (Guardian, 2019b).

While no further details have been forthcoming, the immediate implication for the current Progress 8 secondary school accountability system is clear. If the KS2 tests were scrapped, the adjustment for pupil prior attainment made by Progress 8, or for that matter any other potential school progress measure that Labour might wish to introduce, would be lost (Guardian, 2019a). Thus, assuming no other reforms to school accountability, the scenario where KS2 tests are scrapped would lead to the reduction of Progress 8 to equal Attainment 8, a "raw" or unadjusted measure of the average GCSE exam score in each school. Attainment 8 takes no account of school differences in pupil composition and so the schools with the best results are typically those with the highest prior attaining intakes (Leckie and Goldstein, 2019). The concern if Attainment 8 were to be used in place of Progress 8 is that schools will be rewarded and penalized primarily for who they teach rather than how well they teach. This criticism, which applies to all raw or unadjusted attainment measures of school performance, has been understood by the school effectiveness literature for over 40 years (Mortimore, et al., 1988; Sammons et al, 1997; Rutter et al., 1997; Reynolds et al. 2014; Teddlie and Reynolds, 2000). In contrast, the preferred approach to estimating school effects on pupil attainment is the "value-added" approach (Goldstein, 1997; Raudenbush and Willms, 1995) and Progress 8 is one such example of this approach.

Labour has previously shown acceptance of the need for adjusted measures of school performance. When last in power, Labour introduced Contextual-Value Added (CVA; Ray et al., 2009). CVA, like Progress 8, attempted to measure the average academic progress or improvement pupils make in each school between their end of primary school KS2 tests and their end of secondary school GCSE examinations. CVA, however, additionally accounted for seven pupil background characteristics: age, gender, ethnicity, language, special education needs (SEN), free school meals (FSM), and residential neighbourhood deprivation. The factors adjusted for in CVA had long been known to be predictive of attainment and progress in England (e.g., Thomas, 1998; Strand, 1997). This approach of simultaneously adjusting for both pupil prior attainment and pupil background has long been supported as the preferred approach by the school effectiveness literature (see references above) and this conclusion has been reinforced by several recent studies which compare these competing approaches (Ballou et al., 2004; Lenkeit, 2013; Marks, 2017; Muñoz-Chereau, & Thomas, 2016; Timmermans and Thomas, 2015). CVA, despite its alignment with the academic literature, was removed when the Conservative government replaced Labour after the 2010 general election. See Leckie and Goldstein (2017) for a critical review of the government's arguments for removing CVA.

Given that Progress 8 ignores pupil background, it is no surprise that a number of authors have argued that Progress 8 is biased against schools with educationally disadvantaged intakes (FFT Education Data Lab, 2018; Leckie and Goldstein, 2019; Perry, 2016; TES, 2018). Leckie and Goldstein (2019) illustrated the extent of these biases by contrasting Progress 8 to a new 'Adjusted Progress 8' measure which they propose to reintroduce adjustments for pupil background. They called on the Government to present Adjusted Progress 8 side-by-side with Progress 8 to provide a more informative picture of schools' performances. The Conservative Government, however, are resolute that they will not contextualise Progress 8 in any way (DfE, 2010, 2017, 2019b).

In contrast, given CVA and more generally their education policy history, it seems likely that Labour would prefer to contextualise Attainment 8 by publishing a new 'Adjusted Attainment 8' measure which adjusts for pupil background rather than hold schools to account simply for their raw Attainment 8 scores, which we are considering as the default scenario upon scrapping KS2 tests. This 'contextualised attainment' approach has often been applied in the literature when there is no measure of prior attainment (Lenkeit, 2013). We must stress at this point that while we shall refer to Adjusted Attainment 8 as Labour's measure this is not official labour policy and is just one possible scenario.

Clearly choice of which measure – Attainment 8, Adjusted Attainment 8, Progress 8, or Adjusted Progress 8 – is optimal for measuring school performance and accountability in England is a necessary and timely policy debate. As we have highlighted, these four measures mirror the "raw", "contextualised attainment", "value-added" and "contextual value-added" approaches studied in the literature. While there is substantial agreement as to the relative merits of these competing approaches, what is less clear is the extent to which they will lead to different inferences about secondary schools in England in practice. The aim of this paper is to present the empirical analysis needed to support this policy debate using the government's own data on all pupils and schools in the country. Specifically, we compare Labour's potential Adjusted Attainment 8 measure with the Conservative's Progress 8 measure. This analysis substantially extends that presented in Leckie and Goldstein (2019), which did not consider the possibility that KS2 tests might be scrapped and so did not consider an Adjusted Attainment 8 measure.

**Materials and methods**

We focus on those schools whose Progress 8 scores were published in the Government's 2016 national secondary school performance tables. The data are drawn from the National Pupil Database (gov.uk/government/collections/national-pupil-database) and consist of 502,851 pupils in 3,098 schools. Further details can be found in Leckie and Goldstein (2019).

We calculate all four school performance measures – Attainment 8, Adjusted Attainment 8, Progress 8 and Adjusted Progress 8 – via linear regression, the statistical approach implicit in the Government's own calculation of Progress 8 (DfE, 2019a). Thus, we regress the pupil Attainment 8 score (essentially a total point score across eight GCSE subjects) on an intercept and different sets of pupil-level covariates which vary depending on whether each measure adjusts for pupil prior attainment (KS2 score) or pupil background (age, gender, ethnicity, language, SEN, FSM, and deprivation). These adjustments are summarised in Table 1. Attainment 8 makes no such adjustments and so in this model we enter no covariates. We nevertheless retain the linear regression approach to facilitate comparisons with the other three measures.

The full results for each model can be found in the Supporting Information (Table S1). In each model, the estimated regression coefficients describe the national relationships between Attainment 8 and pupil prior attainment or background. Post-estimation, pupil performance scores are obtained by calculating the pupil residuals from these linear regressions. The school performance scores are then obtained by calculating school averages of their pupil scores. We divide these pupil scores by 10 so that a 1-unit difference on each measure corresponds to a 1-grade difference per subject on the old GCSE A* to G grade scale. The pupil and therefore school scores are calculated after adjusting for the above national relationships. Thus, schools are not held accountable for national relationships, only for how their results deviate from them. Finally, for each school, a 95% confidence interval is calculated to quantify the statistical uncertainty surrounding schools' scores. For each measure, the average score across all schools will be zero. Thus, schools with positive scores and whose 95% confidence intervals lie entirely above zero are described as preforming "significantly above average", while schools with negative scores and whose 95% confidence intervals lie entirely below zero are described as performing "significantly below average".

Associated with each model is also an adjusted R-squared statistic which quantifies the degree to which the estimated national relationships between pupil Attainment 8 and pupil prior attainment and background can predict pupil Attainment 8 scores. The adjusted R-squared statistics for Attainment 8 and Adjusted Progress 8 are 0.00 and 0.62 respectively and represent the extremes associated with making "no adjustments" and making "complete adjustments" for pupil prior attainment and background (complete adjustments at least in so far as the pupil characteristics made available to us in the data). The adjusted R-squared statistics for Adjusted Attainment 8 and Progress 8 are 0.27 and 0.57 and so lie between these two extremes. Thus, these two measures are best described as making only partial adjustments for the different types of pupils taught in each school. Crucially, the low adjusted R-squared for the Adjusted Attainment 8 measure implied by Labour's policy statements suggests that it makes a much weaker adjustment for the different types of pupils taught in different schools than does the current Conservative Progress 8 measure. Indeed, Labour's Adjusted Attainment 8 measure lies closer to the "no adjustment" extreme of Attainment 8 than to the "complete adjustment" extreme of Adjusted Progress 8.

An implication of the Adjusted R-squared statistic increasing across the four models is that the variability of both the pupil and school scores decreases across the four measures (the SDs of the pupil and school scores are presented in Table 1). This makes sense. The more we attribute the variation in pupils' Attainment 8 scores to national trends observed in the data, the less we misattribute the variation to individual pupils and schools.

**Results**

*Comparing the four school performance measures by pupil prior attainment and background characteristics*

Figure 1 presents average pupil scores for each of the four school performance measures by pupil prior attainment and the seven background characteristics.

Thus, Attainment 8, which makes no adjustments for any pupil information at intake shows stark differences in average pupil scores by prior attainment and to a lesser but still important extent by pupil background. Note that we have centred Attainment 8 around the national mean of 5.10 grades per subject on the old GCSE 8 to 1 grade scale. This is in contrast to Government reporting of Attainment 8 which is uncentred and also reported as an aggregate score across the eight subjects. Our reporting of Attainment 8 – in terms of deviations from the national mean and on a per subject basis – is merely to facilitate comparability with the other three measures which all use this approach. For example, pupils who score highest at KS2 (KS2 band 34) are predicted to score 6.15 grades per subject higher than pupils who score lowest at KS2 (KS2 band 1) while FSM pupils score 1.10 grades per subject lower than their more affluent peers. In contrast, the complete adjustment made by Adjusted Progress 8 is confirmed by the zero average pupil scores in every category of prior attainment and each background characteristic.

We see again that Labour's Adjusted Attainment 8 measure and the Conservative's Progress 8 measure lie between these extremes of "no adjustment" and "complete adjustment". Adjusted Attainment 8 adjusts for the seven pupil background characteristics and so shows zero average pupil scores in every category of each background characteristic. However, average pupil scores still vary dramatically across the prior attainment categories and so adjusting for pupil background only goes part way to adjusting the very strong national relationship between Attainment 8 and KS2. Progress 8, in contrast, does adjust for pupil prior attainment, but fails to adjust for pupil background. Progress 8 therefore shows zero average pupil scores in every category of KS2, but average pupil scores which still vary somewhat across the categories of each background characteristic.

In sum, the adjustments made by both Labour's Adjusted Attainment 8 measure and the Conservative's Progress 8 measure appear inadequate. Average pupil scores on Labour's Adjusted Attainment 8 measure vary by KS2 score and so will disadvantage schools with high proportions of low prior attaining pupils. Average pupil scores on the Conservative's Progress 8 measure vary by the pupil background and so will disadvantage schools with, for example, high proportions of FSM and white British pupils. Only the Adjusted Progress 8 measure accounts for the full set of national relationships between pupil Attainment 8 and pupil prior attainment and background.

*Comparing the four school performance measures' scores*

Figure 2 presents scatterplots of schools' scores across the four measures.

The Adjusted Progress 8 against Attainment 8 scatterplot compares the extremes of complete adjustment with no adjustment (top left plot). The Pearson correlation is just 0.62 and so there are many schools with weak Attainment 8 scores who look more impressive when examined through a pure progress lens (schools in the north-west quadrant).

Essentially, these will be schools with very low prior attaining pupils or whose pupils are otherwise especially educationally disadvantaged as measured by their background characteristics. Thus, even though these schools made rapid progress with their pupils, their pupils' Attainment 8 scores are still, on average, low. Similarly, there are many schools with strong Attainment 8 scores who look less convincing once their pupils' prior attainments and backgrounds are taken into account (schools in the south-east quadrant). A case in point is illustrated by the small cluster of schools outlying to the right of the main distribution. These are nearly all grammar schools which, by virtue of entrance examinations, consist of the highest prior attaining pupils and so unsurprisingly they also show the highest Attainment 8 scores some five years later. However, what is interesting is that the average progress made by these pupils, while indeed positive, is not in itself especially impressive.

Turning our attention to Labour's Adjusted Attainment 8 and Conservative's Progress 8 measures (top right plot), the scatterplot shows a correlation of just 0.72 and so these two measures are also measuring different phenomena. To put this correlation in perspective, we note that changing from the Conservative to the Labour measure would lead 1,174 schools (38% of all schools in the country) to move up or down the national league table of all schools by 500 or more places, with 364 schools (12%) moving over 1,000 places. Thus, many schools are viewed quite differently depending on which measure is used. The outlying cluster of schools positioned at the top of the plot are again nearly all grammar schools. This positioning indicates that Labour's Adjusted Attainment 8 measure would look on grammar schools far more favourably than is the case for the current Conservative Progress 8 measure.

More worryingly still, the scatterplots of Labour's Adjusted Attainment 8 measure against the "no adjustments" extreme of Attainment 8 (middle left plot) and the "complete adjustments" extreme of Adjusted Progress 8 (middle right plot) show correlations of 0.87 and 0.75. This is in contrast to the scatterplots of Conservatives' Progress 8 measure against the "no adjustments" extreme of Attainment 8 (bottom left plot) and the "complete adjustments" extreme of Adjusted Progress 8 (bottom right plot) which show correlations of 0.75 and 0.91. This ordering of correlations is consistent with the Adjusted R-squared statistics; Labour and Conservative's measures only make partial adjustments for pupil prior attainment and background. Crucially, as Labour's Adjusted Attainment 8 measure lies closer to the "no adjustment" Attainment 8 measure than the "complete adjustment" Adjusted Progress 8 measure (0.87 > 0.75) this suggests that Labour's measure is a less meaningful and fair measure of school performance than is the Conservative's measure. However, we argue that the preferred measure continues to be Adjusted Progress 8 which simultaneously adjusts for pupil prior attainment and pupil background simultaneously.

*Comparing the four school performance measures' scores by school characteristics*

Figure 3 presents average pupil scores for each of the four school performance measures by seven school characteristics: region, type, admissions policy, age range, gender, religious denomination, and neighbourhood deprivation.

Attainment 8, which makes no adjustments for any pupil information at intake, shows the starkest differences in average pupil scores by school characteristics. We see especially strong differentials by school type with, for example, pupils in converter academies scoring 0.71 grades per subject higher than pupils in sponsored academies, and by school admissions with grammar school pupils scoring 2.11 grades higher per subject than pupils in secondary moderns. However, little should be read into these statistics regarding potential quality differences in education as the pupils attending these different categories of schools are so very different. In contrast, the Adjusted Progress 8 measure, which attempts to account for differences in both pupil prior attainment and background, shows far smaller differentials by school characteristics. However, these differentials, while reduced, are not trivial and importantly cannot simply be explained by the different types of pupil attending different school types. They are therefore a cause for concern and further investigation. There are some quite specific concerns, for example, the lower performance of pupils in UTCs and Studio schools is likely to reflect the different curriculum often pursed in these schools. Furthermore, the fact that these schools typically start educating at 14 raises the question as to whether it is appropriate at all to be comparing these schools to the vast majority which teach from age 11 and for which adjusting for KS2 scores as a measure of prior attainment is more intuitive.

The results for Labour's Adjusted Attainment 8 measure and the Conservative's Progress 8 measure once again lie between the extremes presented by "no adjustment" Attainment 8 and "complete adjustment" Adjusted Progress 8. We can see in Figure 3 that the variability of the differentials in general reduces as we more from left to right across the four measures. This dampening down of the differentials is consistent with the increasing Adjusted R-squared statistics across the four models and the decreasing SDs for pupil scores.

We see again that Labour's Adjusted Attainment 8 measure views Grammar schools far more favourably than does Conservative's Progress 8 measure, but, as explained earlier, this difference reflects the inadequate adjustment that Labour's Adjusted Attainment 8 makes for pupil prior attainment.

There are some interesting instances where the reported differentials do not decrease monotonically as we move across the four measures, despite the increasing explanatory power of the models. In terms of the nine regions, for example, the North East is ranked seventh on Attainment 8, reflecting the relatively low average GCSE results seen in this region. However, the North East, is ranked second after adjustment for the seven pupil background characteristics. This improvement in its relative standing primarily reflects the above average proportions of FSM and white British pupils schooled in this region, two pupil groups which struggle nationally in terms of Attainment 8. Shifting to Progress 8, pupil background is once again ignored, and it is pupil prior attainment instead which is adjusted. As a result, the North East is once again ranked seventh indicating that whether we focus on average final attainment or average improvement in attainment, the region appears to be struggling compared to most other regions. However, moving to Adjusted Progress 8 the story changes again. The region is now ranked third, again suggesting that once we acknowledge that the pupils taught in the North East are somewhat more educationally disadvantaged than pupils nationally, the average progress shown by pupils in this region should also be viewed more favourably.

**Discussion**

In this article, we have argued that Labour's announcement to scrap national testing at KS2 has immediate important ramifications for the Progress 8 secondary school accountability system. In absence of any further reforms, scrapping the KS2 tests would likely result in a new headline school performance measure which would prove less fair and meaningful for comparing schools than the Conservative's Progress 8 measure which itself has been widely criticised as being biased against schools teaching educationally disadvantaged intakes. Indeed, our empirical analyses shows that in a potential accountability system that removes the primary source of prior attainment information, the likely best-case scenario of a new 'Adjusted Attainment 8' measure, would lie closer to Attainment 8, a simple comparison of school average exam results, than it would to the 'Adjusted Progress 8' measure proposed by Leckie and Goldstein (2019) and which was argued there to be the most appropriate measure given the data. Thus, simply scrapping the KS2 tests, in order to address the many valid concerns of educationalists and parents, would likely create new problems for the current secondary school accountability system which do not appear to have been adequately thought through by Labour.

Our conclusions may be viewed as especially worrying given that the underlying arguments we make and our empirical findings are entirely consistent with the substantial and long-standing literature on measuring school performance (Goldstein, 1997; Mortimore, et al., 1988; Sammons et al, 1997; Raudenbush and Willms, 1995; Rutter et al., 1997; Reynolds et al. 2014; Teddlie and Reynolds, 2000) and specifically on comparing "raw", "value-added", "contextual attainment", and "contextual value-added" models (Ballou et al., 2004; Lenkeit, 2013; Marks, 2017; Muñoz-Chereau, & Thomas, 2016; Timmermans and Thomas, 2015). That is, we show, that adjusting for prior attainment and pupil background (demographic and socioeconomic characteristics) lead to fundamental changes in how schools are evaluated and that the preferred approach is to simultaneously adjust for both sets of factors. The importance of our work is not that we replicate this established finding, it is that we draw attention to and start a much-needed debate around the fact that both the Labour and Conservative political parties appear to ignore this finding. The policy choice between Attainment 8, Adjusted Attainment 8, Progress 8 and Adjusted Progress 8 matters and does not appear to be informed by the substantial research evidence on measuring school performance.

In this article we have focussed on the immediate implications of scrapping KS2 tests on Progress 8 and secondary school accountability. At the primary school level, the scrapping of the national KS2 tests proposed by Labour would necessitate an abandoning of the current school performance tables altogether; there would be no measure of pupil attainment at the end of primary schooling by which schools could be compared. It may be the case that Labour ultimately wishes to also scrap secondary school performances measures. However, at secondary level, GCSE examinations are statutory and there is no suggestion that Labour would additionally scrap these assessments. Furthermore, given that schools' GCSE results have been published since their inception it seems likely that they will continue to appear in the public domain, if nothing else, via freedom of information requests made by newspapers and others. Given this, it would seem that Labour would continue to publish official secondary school performance tables rather than lose the narrative around how schools are performing to the media and others who may present the data in a less careful and sensitive manner. One approach that Labour might then take would be to declare that schools will no longer be held directly accountable for their scores and that the scores are only published on a public interest basis. However, the presence of the scores in the public domain would always lead schools to be held publicly accountable with, for example, parents shunning schools which appear to do badly. Thus, pressures would remain on schools to perform well on whatever scores are published in the tables. Scrapping KS2 tests will not lessen these pressures, rather it will make matters more problematic as schools will be judged instead by less meaningful measures. One might hope that the obvious inadequacy of unadjusted Attainment 8 GCSE scores would persuade people not to rely on them for school comparisons. However, this is not something which ever transpired over the 25-year period that 5+A*-C (the percentage of pupils getting five or more GCSEs at grade C or above) was the headline school performance measure (Leckie and Goldstein, 2017).

To the extent to which Labour might declare that the school performance tables will no longer play a direct role in the secondary school accountability system, the default way forward would then be to rely solely on Ofsted, the national school inspectorate. Here too, however, Labour plan change. At their September 2019 annual party conference, Angela Rayner, Labour Shadow Education Secretary announced (Labour, 2019b):

> Labour's [*sic*] will replace Ofsted with a two-phase inspection system – all schools and education providers will be subject to regular 'health checks' led by local government, and a more in-depth inspection led by Her Majesty's Inspectors (HMIs) […] HMIs will carry out inspections in response to concerns arising from these the regular 'health checks'

An important criticism of Ofsted has long been that it relies too heavily on school performance table data when awarding their ratings (NAHT, 2018). Thus, a deemphasis or removal of the official tables would force Ofsted or its replacement to rely more on what they learn from the school inspection visits themselves. A possible concern here is the subjective nature of school inspectors' ratings and judgements (YouGov, 2018), although many would argue that this is inevitable and that well-trained human judges are desirable. Furthermore, given the likely continuation of school performance data in the media, there would be concerns as to just how independently Ofsted or its replacement would make their ratings and judgements, especially for those schools who appear weak in terms of these data.

Thus, it would seem hard to abolish school performance tables entirely and therefore the inevitable assessment of schools via the GCSE examination results of the children themselves. As the literature has long shown and our work has demonstrated yet again, it is important that school GCSE examination results are sufficiently contextualised by adjusting for both pupil prior attainment and background. However, this necessitates the retention of the KS2 tests and, it would seem, the problems associated with them. One way out of this conundrum might be to modify the way the KS2 tests are implemented and used in such a way as to reduce the concerns around them. Perhaps the most obvious conclusion is to retain the KS2 tests, but to no longer publish results on a primary school by primary school basis, but still use the KS2 tests in their adjustment role for secondary school performance and accountability measures. Such an approach would be analogous to the way KS1 data are currently not published but are used in primary school accountability by giving a KS1 to KS2 progress measure. By no longer publishing the KS2 tests and only using them as an input for secondary school accountability this would reduce the high stakes currently associated with the KS2 tests. This in turn would give primary schools the freedom to educate away from the pressure to teach to the test and the narrowing of the curriculum that the current system has led too.

Table 1.

Overview of Attainment 8, Adjusted Attainment 8, Progress 8 and Adjusted Progress 8.

|  | School performance measure | | | |
| --- | --- | --- | --- | --- |
|  | Attainment 8 | Adjusted Attainment 8 | Progress 8 | Adjusted Progress 8 |
| Supported by |  | Labour | Conservative |  |
| Statistical adjustments | None | Partial | Partial | Complete |
| Prior attainment | No | No | Yes | Yes |
| Pupil background | No | Yes | No | Yes |
| Adjusted R-squared | 0.00 | 0.27 | 0.57 | 0.62 |
| SD of pupil scores | 1.62 | 1.38 | 1.06 | 0.99 |
| SD of school scores | 0.75 | 0.53 | 0.40 | 0.35 |

Note.

Sample size = 502,851 pupils in 3,098 schools.

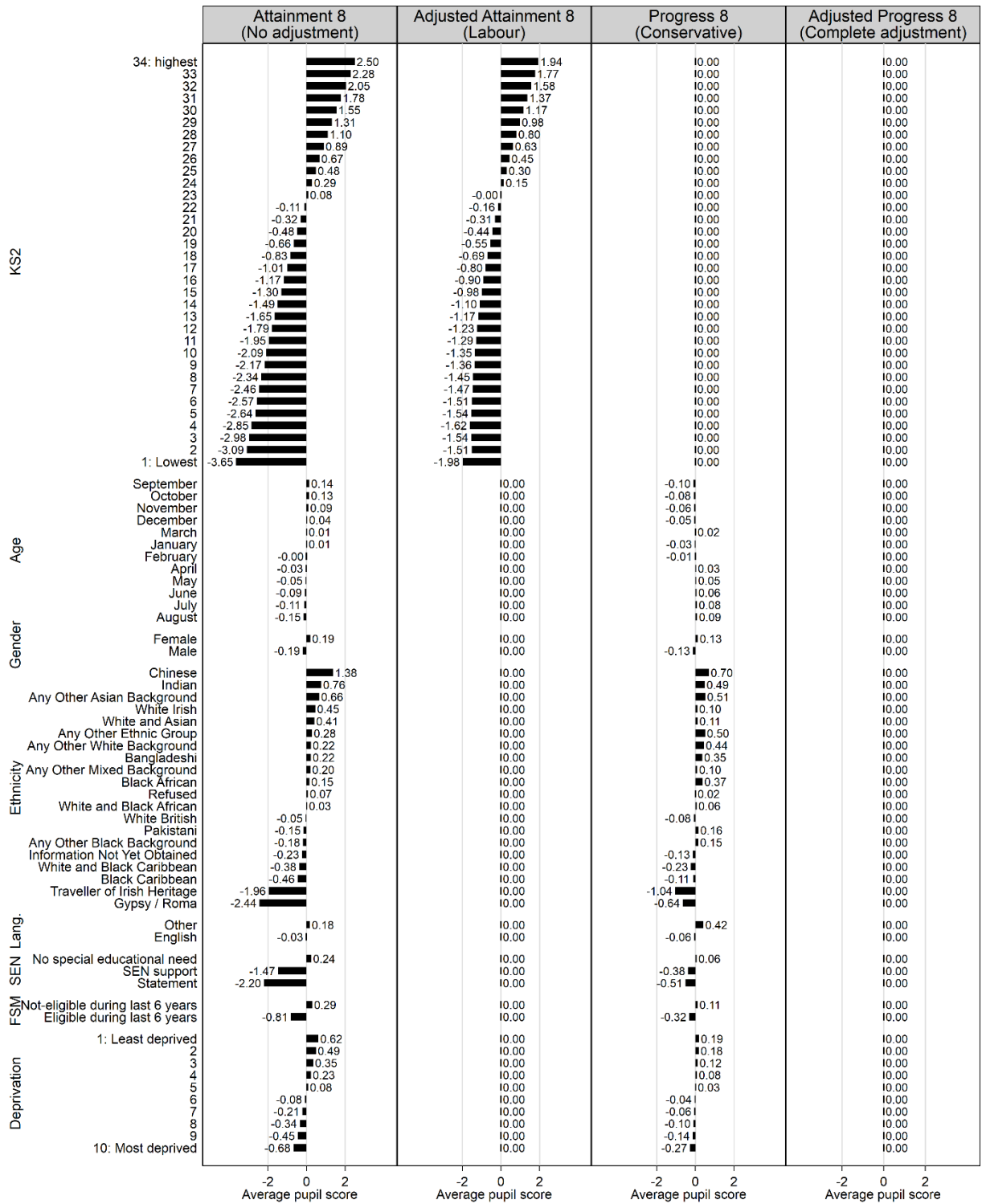

Figure 1.

Average Attainment 8, Adjusted Attainment 8, Progress 8 and Adjusted Progress 8 scores by pupil characteristics.

Note. The number of pupils by pupil characteristic are given in the Supporting Information (Table S2).

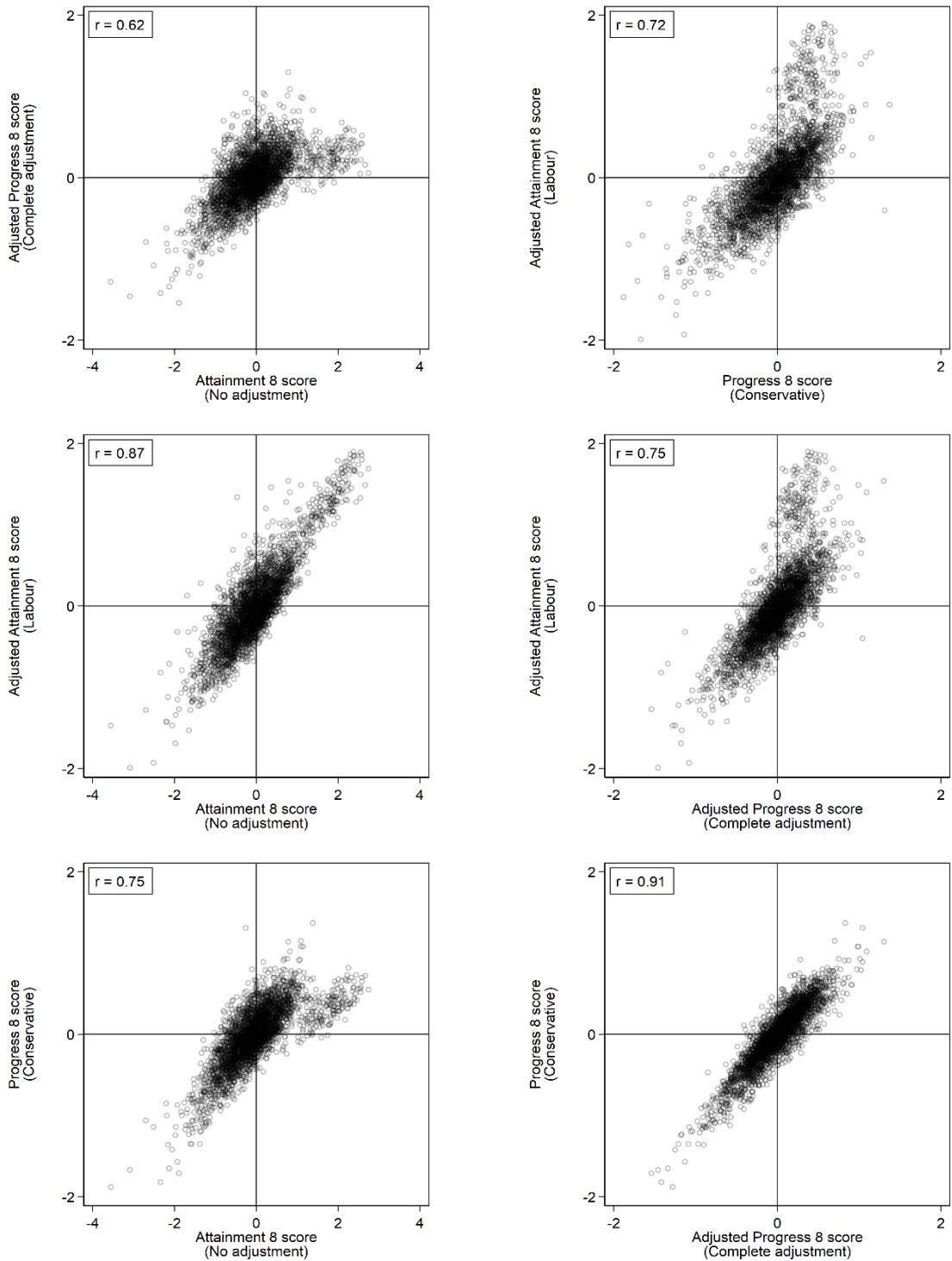

Figure 2.

Scatterplots of school average Attainment 8, Adjusted Attainment 8, Progress 8 and Adjusted Progress 8 scores with Pearson correlations.

Note.

The horizontal and vertical lines denote the mean values of the relevant variables.

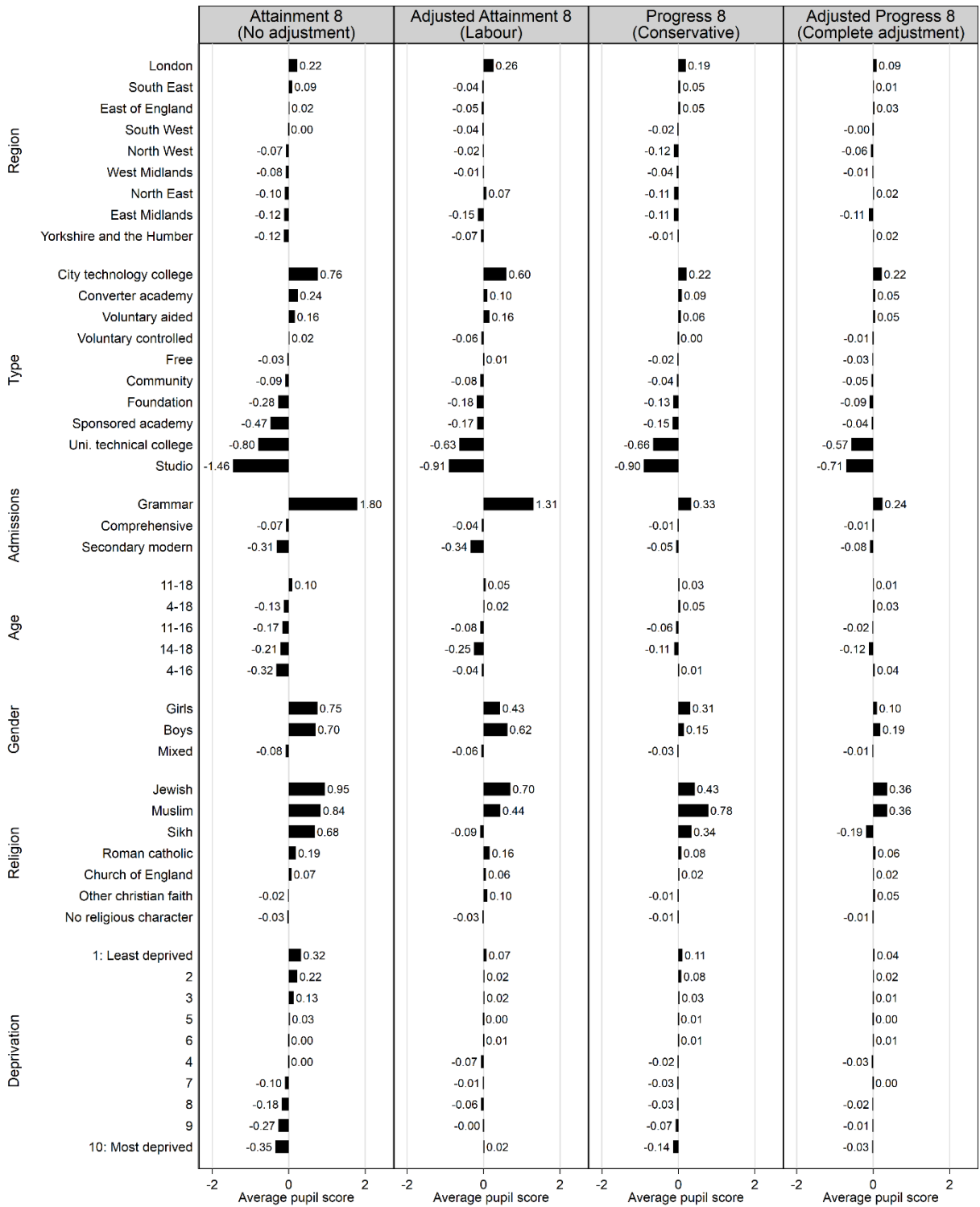

Figure 3.

Average Attainment 8, Adjusted Attainment 8, Progress 8 and Adjusted Progress 8 scores by school characteristics.

Note.

The categories of each school characteristic are sorted by Attainment 8 score.

The number of schools by school characteristic is given in the Supporting Information (Table S3).

**Supplemental materials**

Table S1.

Model results for Attainment 8, Adjusted Attainment 8, Progress 8 and Adjusted Progress 8 linear regression models.

| Variable | Attainment 8 | | Adjusted Attainment 8 | | Progress 8 | | Adjusted Progress 8 | |
|---|---|---|---|---|---|---|---|---|
| | Coef. | SE | Coef. | SE | Coef. | SE | Coef. | SE |
| Constant | 51.02* | 0.11 | 58.61* | 0.17 | 14.52* | 0.55 | 19.74* | 0.51 |
| KS2 group (ref. cat. = KS2 Group 1) | | | | | | | | |
| KS2 group 2 | | | | | 5.55* | 0.66 | 5.52* | 0.59 |
| KS2 group 3 | | | | | 6.73* | 0.54 | 6.73* | 0.48 |
| KS2 group 4 | | | | | 8.00* | 0.57 | 7.71* | 0.52 |
| KS2 group 5 | | | | | 10.11* | 0.60 | 9.29* | 0.54 |
| KS2 group 6 | | | | | 10.83* | 0.60 | 9.86* | 0.55 |
| KS2 group 7 | | | | | 11.94* | 0.57 | 10.84* | 0.52 |
| KS2 group 8 | | | | | 13.11* | 0.58 | 11.67* | 0.53 |
| KS2 group 9 | | | | | 14.78* | 0.57 | 13.04* | 0.52 |
| KS2 group 10 | | | | | 15.62* | 0.57 | 13.63* | 0.52 |
| KS2 group 11 | | | | | 16.97* | 0.56 | 14.75* | 0.51 |
| KS2 group 12 | | | | | 18.62* | 0.56 | 16.03* | 0.52 |
| KS2 group 13 | | | | | 20.04* | 0.56 | 17.22* | 0.52 |
| KS2 group 14 | | | | | 21.56* | 0.56 | 18.48* | 0.51 |

| | | | | | |
|---|---|---|---|---|---|
| KS2 group 15 | | | 23.47* | 0.56 | 20.09* | 0.51 |
| KS2 group 16 | | | 24.83* | 0.56 | 21.24* | 0.51 |
| KS2 group 17 | | | 26.43* | 0.55 | 22.72* | 0.51 |
| KS2 group 18 | | | 28.16* | 0.55 | 24.18* | 0.51 |
| KS2 group 19 | | | 29.94* | 0.55 | 25.86* | 0.51 |
| KS2 group 20 | | | 31.70* | 0.55 | 27.38* | 0.51 |
| KS2 group 21 | | | 33.34* | 0.55 | 28.89* | 0.51 |
| KS2 group 22 | | | 35.43* | 0.55 | 30.76* | 0.51 |
| KS2 group 23 | | | 37.33* | 0.55 | 32.53* | 0.51 |
| KS2 group 24 | | | 39.39* | 0.55 | 34.40* | 0.51 |
| KS2 group 25 | | | 41.32* | 0.55 | 36.18* | 0.51 |
| KS2 group 26 | | | 43.17* | 0.55 | 37.87* | 0.51 |
| KS2 group 27 | | | 45.40* | 0.55 | 39.94* | 0.51 |
| KS2 group 28 | | | 47.51* | 0.55 | 41.92* | 0.51 |
| KS2 group 29 | | | 49.62* | 0.55 | 43.93* | 0.51 |
| KS2 group 30 | | | 52.01* | 0.55 | 46.11* | 0.51 |
| KS2 group 31 | | | 54.30* | 0.56 | 48.27* | 0.51 |
| KS2 group 32 | | | 56.96* | 0.56 | 50.69* | 0.51 |
| KS2 group 33 | | | 59.34* | 0.56 | 52.90* | 0.52 |
| KS2 group 34 | | | 61.54* | 0.56 | 54.91* | 0.52 |
| Month of birth (ref. cat. = September) | | | | | | |
| October | -0.01 | 0.10 | | | 0.15* | 0.07 |
| November | -0.33* | 0.10 | | | 0.35* | 0.07 |
| December | -0.68* | 0.10 | | | 0.42* | 0.07 |

| | | | | |
|---|---:|---:|---:|---:|
| January | -0.94* | 0.10 | 0.59* | 0.07 |
| February | -1.02* | 0.10 | 0.78* | 0.07 |
| March | -1.02* | 0.10 | 0.99* | 0.07 |
| April | -1.32* | 0.10 | 1.12* | 0.07 |
| May | -1.59* | 0.10 | 1.21* | 0.07 |
| June | -1.85* | 0.10 | 1.30* | 0.07 |
| July | -1.94* | 0.09 | 1.49* | 0.07 |
| August | -2.14* | 0.10 | 1.62* | 0.07 |
| Gender (ref. cat. = Male) | | | | |
| Female | 2.95* | 0.10 | 2.44* | 0.05 |
| Ethnicity (ref. cat. = White British) | | | | |
| White Irish | 4.81* | 0.40 | 2.02* | 0.29 |
| Traveller of Irish Heritage | -10.36* | 2.18 | -6.92* | 1.63 |
| Gypsy / Roma | -16.06* | 0.92 | -5.63* | 0.68 |
| Any Other White Background | 3.95* | 0.20 | 3.90* | 0.13 |
| Black African | 6.82* | 0.21 | 5.42* | 0.16 |
| Black Caribbean | 1.10* | 0.26 | 1.80* | 0.20 |
| Any Other Black Background | 3.65* | 0.33 | 3.75* | 0.25 |
| Indian | 7.49* | 0.30 | 4.16* | 0.18 |
| Pakistani | 2.11* | 0.30 | 1.93* | 0.19 |
| Bangladeshi | 7.07* | 0.34 | 4.49* | 0.22 |
| Any Other Asian Background | 7.47* | 0.34 | 4.71* | 0.16 |
| Chinese | 12.96* | 0.39 | 6.26* | 0.22 |
| White and Black African | 3.87* | 0.30 | 2.46* | 0.23 |
| White and Black Caribbean | 0.27 | 0.20 | 0.04 | 0.15 |
| White and Asian | 4.68* | 0.23 | 2.08* | 0.16 |
| Any Other Mixed Background | 4.26* | 0.23 | 2.32* | 0.15 |
| Any Other Ethnic Group | 7.14* | 0.29 | 5.67* | 0.18 |
| Information Not Yet Obtained | -0.36 | 1.27 | -0.14 | 0.69 |
| Refused | 2.45* | 0.36 | 1.36* | 0.26 |
| First language (ref. cat. English) | | | | |
| Other | 0.17 | 0.16 | 2.55* | 0.10 |
| SEN (ref. cat. = None) | | | | |
| SEN support | -14.39* | 0.13 | -4.42* | 0.10 |
| Statement | -22.45* | 0.30 | -6.88* | 0.33 |
| Eligible for FSM (ref. cat. = No) | | | | |
| Yes | -7.74* | 0.08 | -4.01* | 0.05 |
| Deprivation (ref. cat. = decile 1) | | | | |
| IDACI decile 2 | -1.34* | 0.13 | -0.22* | 0.09 |
| IDACI decile 3 | -2.48* | 0.14 | -0.79* | 0.10 |
| IDACI decile 4 | -3.56* | 0.15 | -1.28* | 0.10 |
| IDACI decile 5 | -4.65* | 0.15 | -1.87* | 0.11 |
| IDACI decile 6 | -5.98* | 0.16 | -2.66* | 0.11 |
| IDACI decile 7 | -6.81* | 0.17 | -2.99* | 0.12 |
| IDACI decile 8 | -7.62* | 0.17 | -3.43* | 0.12 |
| IDACI decile 9 | -8.13* | 0.18 | -3.82* | 0.13 |
| IDACI decile 10 | -9.06* | 0.20 | -4.52* | 0.15 |
| Adjusted R-squared | 0.000 | 0.270 | 0.570 | 0.624 |

Note.

Coef. = Regression coefficient.

* = $p < 0.05$ (standard errors and statistical tests adjust for school-level clustering).

Sample size = 502,851 pupils in 3,098 schools.

Table S2.

Distribution of pupils and pupil average Attainment 8, Adjusted Attainment 8, Progress 8 and Adjusted Progress 8 scores by pupil demographic and socioeconomic characteristics.

| Variable | Pupils | | A8 | AA8 | P8 | AP8 |
|---|---|---|---|---|---|---|
| | N | % | | | | |
| KS2 prior attainment group | | | | | | |
| 1: Lowest | 960 | 0.2 | -3.65* | -1.98* | 0.00 | 0.00 |
| 2 | 1164 | 0.2 | -3.09* | -1.51* | 0.00 | 0.00 |
| 3 | 7692 | 1.5 | -2.98* | -1.54* | 0.00 | 0.00 |
| 4 | 3133 | 0.6 | -2.85* | -1.62* | 0.00 | 0.00 |
| 5 | 2413 | 0.5 | -2.64* | -1.54* | 0.00 | 0.00 |
| 6 | 2417 | 0.5 | -2.57* | -1.51* | 0.00 | 0.00 |
| 7 | 3287 | 0.7 | -2.46* | -1.47* | 0.00 | 0.00 |
| 8 | 3359 | 0.7 | -2.34* | -1.45* | 0.00 | 0.00 |
| 9 | 4757 | 0.9 | -2.17* | -1.36* | 0.00 | 0.00 |
| 10 | 5228 | 1.0 | -2.09* | -1.35* | 0.00 | 0.00 |
| 11 | 6357 | 1.3 | -1.95* | -1.29* | 0.00 | 0.00 |
| 12 | 7499 | 1.5 | -1.79* | -1.23* | 0.00 | 0.00 |
| 13 | 8337 | 1.7 | -1.65* | -1.17* | 0.00 | 0.00 |
| 14 | 10041 | 2.0 | -1.49* | -1.10* | 0.00 | 0.00 |
| 15 | 12033 | 2.4 | -1.30* | -0.98* | 0.00 | 0.00 |
| 16 | 13679 | 2.7 | -1.17* | -0.90* | 0.00 | 0.00 |
| 17 | 16026 | 3.2 | -1.01* | -0.80* | 0.00 | 0.00 |
| 18 | 19589 | 3.9 | -0.83* | -0.69* | 0.00 | 0.00 |
| 19 | 23473 | 4.7 | -0.66* | -0.55* | 0.00 | 0.00 |
| 20 | 25852 | 5.1 | -0.48* | -0.44* | 0.00 | 0.00 |
| 21 | 29549 | 5.9 | -0.32* | -0.31* | 0.00 | 0.00 |
| 22 | 30450 | 6.1 | -0.11* | -0.16* | 0.00 | 0.00 |
| 23 | 30669 | 6.1 | 0.08* | 0.00 | 0.00 | 0.00 |
| 24 | 31371 | 6.2 | 0.29* | 0.15* | 0.00 | 0.00 |
| 25 | 30990 | 6.2 | 0.48* | 0.30* | 0.00 | 0.00 |
| 26 | 29952 | 6.0 | 0.67* | 0.45* | 0.00 | 0.00 |
| 27 | 28983 | 5.8 | 0.89* | 0.63* | 0.00 | 0.00 |
| 28 | 27346 | 5.4 | 1.10* | 0.80* | 0.00 | 0.00 |
| 29 | 24938 | 5.0 | 1.31* | 0.98* | 0.00 | 0.00 |
| 30 | 21913 | 4.4 | 1.55* | 1.17* | 0.00 | 0.00 |
| 31 | 18167 | 3.6 | 1.78* | 1.37* | 0.00 | 0.00 |
| 32 | 12225 | 2.4 | 2.05* | 1.58* | 0.00 | 0.00 |
| 33 | 6505 | 1.3 | 2.28* | 1.77* | 0.00 | 0.00 |
| 34: Highest | 2497 | 0.5 | 2.50* | 1.94* | 0.00 | 0.00 |
| Month of birth | | | | | | |
| September | 43346 | 8.6 | 0.14* | 0.00 | -0.10* | 0.00 |
| October | 41981 | 8.3 | 0.13* | 0.00 | -0.08* | 0.00 |
| November | 41113 | 8.2 | 0.09* | 0.00 | -0.06* | 0.00 |
| December | 42700 | 8.5 | 0.04* | 0.00 | -0.05* | 0.00 |
| January | 42124 | 8.4 | 0.01 | 0.00 | -0.03* | 0.00 |
| February | 38949 | 7.7 | 0.00 | 0.00 | -0.01 | 0.00 |
| March | 42158 | 8.4 | 0.01 | 0.00 | 0.02* | 0.00 |
| April | 40458 | 8.0 | -0.03* | 0.00 | 0.03* | 0.00 |
| May | 42601 | 8.5 | -0.05* | 0.00 | 0.05* | 0.00 |
| June | 40983 | 8.2 | -0.09* | 0.00 | 0.06* | 0.00 |
| July | 43493 | 8.6 | -0.11* | 0.00 | 0.08* | 0.00 |
| August | 42945 | 8.5 | -0.15* | 0.00 | 0.09* | 0.00 |
| Gender | | | | | | |
| Male | 253733 | 50.5 | -0.19* | 0.00 | -0.13* | 0.00 |
| Female | 249118 | 49.5 | 0.19* | 0.00 | 0.13* | 0.00 |
| Ethnicity | | | | | | |
| White British | 380949 | 75.8 | -0.05* | 0.00 | -0.08* | 0.00 |
| White Irish | 1606 | 0.3 | 0.45* | 0.00 | 0.10* | 0.00 |
| Traveller of Irish Heritage | 104 | 0.0 | -1.96* | 0.00 | -1.04* | 0.00 |
| Gypsy / Roma | 659 | 0.1 | -2.44* | 0.00 | -0.64* | 0.00 |
| Any Other White Background | 17129 | 3.4 | 0.22* | 0.00 | 0.44* | 0.00 |
| Black African | 14379 | 2.9 | 0.15* | 0.00 | 0.37* | 0.00 |
| Black Caribbean | 6650 | 1.3 | -0.46* | 0.00 | -0.11* | 0.00 |
| Any Other Black Background | 2690 | 0.5 | -0.18* | 0.00 | 0.15* | 0.00 |
| Indian | 12426 | 2.5 | 0.76* | 0.00 | 0.49* | 0.00 |
| Pakistani | 18722 | 3.7 | -0.15* | 0.00 | 0.16* | 0.00 |
| Bangladeshi | 7709 | 1.5 | 0.22* | 0.00 | 0.35* | 0.00 |
| Any Other Asian Background | 6900 | 1.4 | 0.66* | 0.00 | 0.51* | 0.00 |
| Chinese | 1585 | 0.3 | 1.38* | 0.00 | 0.70* | 0.00 |
| White and Black African | 2390 | 0.5 | 0.03 | 0.00 | 0.06* | 0.00 |
| White and Black Caribbean | 6873 | 1.4 | -0.38* | 0.00 | -0.23* | 0.00 |
| White and Asian | 4656 | 0.9 | 0.41* | 0.00 | 0.11* | 0.00 |
| Any Other Mixed Background | 6983 | 1.4 | 0.20* | 0.00 | 0.10* | 0.00 |
| Any Other Ethnic Group | 6198 | 1.2 | 0.28* | 0.00 | 0.50* | 0.00 |
| Information Not Yet Obtained | 2098 | 0.4 | -0.23 | 0.00 | -0.13 | 0.00 |
| Refused | 2145 | 0.4 | 0.07 | 0.00 | 0.02 | 0.00 |
| English as additional language | | | | | | |
| English as first language | 438585 | 87.2 | -0.03* | 0.00 | -0.06* | 0.00 |
| English as additional language | 64266 | 12.8 | 0.18* | 0.00 | 0.42* | 0.00 |
| Special educational needs | | | | | | |
| No special educational need | 436229 | 86.8 | 0.24* | 0.00 | 0.06* | 0.00 |
| SEN support | 55601 | 11.1 | -1.47* | 0.00 | -0.38* | 0.00 |
| Statement | 11021 | 2.2 | -2.20* | 0.00 | -0.51* | 0.00 |
| Free school meal status | | | | | | |
| Not-eligible during last 6 years | 369147 | 73.4 | 0.29* | 0.00 | 0.11* | 0.00 |
| Eligible during last 6 years | 133704 | 26.6 | -0.81* | 0.00 | -0.32* | 0.00 |
| Deprivation | | | | | | |
| 1: Least deprived | 50289 | 10.0 | 0.62* | 0.00 | 0.19* | 0.00 |
| 2 | 51790 | 10.3 | 0.49* | 0.00 | 0.18* | 0.00 |
| 3 | 49086 | 9.8 | 0.35* | 0.00 | 0.12* | 0.00 |
| 4 | 51072 | 10.2 | 0.23* | 0.00 | 0.08* | 0.00 |
| 5 | 50340 | 10.0 | 0.08* | 0.00 | 0.03* | 0.00 |
| 6 | 49321 | 9.8 | -0.08* | 0.00 | -0.04* | 0.00 |
| 7 | 50172 | 10.0 | -0.21* | 0.00 | -0.06* | 0.00 |
| 8 | 50853 | 10.1 | -0.34* | 0.00 | -0.10* | 0.00 |
| 9 | 49761 | 9.9 | -0.45* | 0.00 | -0.14* | 0.00 |
| 10: Most deprived | 50167 | 10.0 | -0.68* | 0.00 | -0.27* | 0.00 |

Note.

Sample size = 502,851 pupils in 3,098 schools.

A8 = Attainment 8.

AA8 = Adjusted Attainment 8.

P8 = Progress 8.

AP8 = Adjusted Progress 8.

* = $p < 0.05$ (statistical tests adjust for school-level clustering).

Table S3.

Distribution of pupils and schools and pupil average Attainment 8, Adjusted Attainment 8, Progress 8 and Adjusted Progress 8 scores by school characteristics.

| Characteristic | Schools | | A8 | AA8 | P8 | AP8 |
|---|---|---|---|---|---|---|
| | N | % | | | | |
| Region | | | | | | |
| London | 431 | 13.9 | 0.22* | 0.26* | 0.19* | 0.09* |
| South East | 474 | 15.3 | 0.09* | -0.04 | 0.05* | 0.01 |
| South West | 309 | 10.0 | 0.00 | -0.04 | -0.02 | -0.00 |
| West Midlands | 373 | 12.0 | -0.08* | -0.01 | -0.04* | -0.01 |
| North West | 447 | 14.4 | -0.07* | -0.02 | -0.12* | -0.06* |
| North East | 152 | 4.9 | -0.10* | 0.07* | -0.11* | 0.02 |
| Yorkshire & Humber | 298 | 9.6 | -0.12* | -0.07* | -0.01 | 0.02 |
| East Midlands | 269 | 8.7 | -0.12* | -0.15* | -0.11* | -0.11* |
| East of England | 345 | 11.1 | 0.02 | -0.05* | 0.05* | 0.03 |
| School type | | | | | | |
| Community | 538 | 17.4 | -0.09* | -0.08* | -0.04* | -0.05* |
| Foundation | 275 | 8.9 | -0.28* | -0.18* | -0.13* | -0.09* |
| Voluntary aided | 273 | 8.8 | 0.16* | 0.16* | 0.06* | 0.05* |
| Voluntary controlled | 34 | 1.1 | 0.02 | -0.06 | 0.00 | -0.01 |
| City tech. college | 3 | 0.1 | 0.76* | 0.60* | 0.22 | 0.22 |
| Sponsored academy | 560 | 18.1 | -0.47* | -0.17* | -0.15* | -0.04* |
| Converter academy | 1320 | 42.6 | 0.24* | 0.10* | 0.09* | 0.05* |
| Free | 27 | 0.9 | -0.03 | 0.01 | -0.02 | -0.03 |
| Studio | 30 | 1.0 | -1.46* | -0.91* | -0.90* | -0.71* |
| Uni. tech. college | 26 | 0.8 | -0.80* | -0.63* | -0.66* | -0.57* |
| Further ed. college | 12 | 0.4 | -2.58* | -0.93* | -1.82* | -1.39* |
| School admissions | | | | | | |
| Comprehensive | 2819 | 91.0 | -0.07* | -0.04* | -0.01 | -0.01 |
| Grammar | 162 | 5.2 | 1.80* | 1.31* | 0.33* | 0.24* |
| Secondary modern | 117 | 3.8 | -0.31* | -0.34* | -0.05 | -0.08* |
| Age range | | | | | | |
| 11-18 | 1881 | 60.7 | 0.10* | 0.05* | 0.03* | 0.01* |
| 11-16 | 971 | 31.3 | -0.17* | -0.08* | -0.06* | -0.02 |
| 14-18 | 135 | 4.4 | -0.21* | -0.25* | -0.11* | -0.12* |
| 4-18 | 83 | 2.7 | -0.13 | 0.02 | 0.05 | 0.03 |
| 4-16 | 28 | 0.9 | -0.32* | -0.04 | 0.01 | 0.04 |
| School gender | | | | | | |
| Mixed | 2738 | 88.4 | -0.08* | -0.06* | -0.03* | -0.01* |
| Boys | 151 | 4.9 | 0.70* | 0.62* | 0.15* | 0.19* |
| Girls | 209 | 6.7 | 0.75* | 0.43* | 0.31* | 0.10* |
| School religion | | | | | | |
| None | 2524 | 81.5 | -0.03* | -0.03* | -0.01 | -0.01 |
| Church of England | 176 | 5.7 | 0.07 | 0.06 | 0.02 | 0.02 |
| Roman catholic | 310 | 10.0 | 0.19* | 0.16* | 0.08* | 0.06* |
| Other Christian faith | 68 | 2.2 | -0.02 | 0.10 | -0.01 | 0.05 |
| Jewish | 11 | 0.4 | 0.95* | 0.70* | 0.43* | 0.36* |
| Muslim | 8 | 0.3 | 0.84* | 0.44* | 0.78* | 0.36* |
| Sikh | 1 | 0.0 | 0.68* | -0.09* | 0.34* | -0.19* |
| School IDACI decile | | | | | | |
| 1: Least deprived | 288 | 9.3 | 0.32* | 0.07* | 0.11* | 0.04* |
| 2 | 329 | 10.6 | 0.22* | 0.02 | 0.08* | 0.02 |
| 3 | 313 | 10.1 | 0.13* | 0.02 | 0.03 | 0.01 |
| 4 | 303 | 9.8 | 0.00 | -0.07* | -0.02 | -0.03* |
| 5 | 325 | 10.5 | 0.03 | 0.00 | 0.01 | 0.00 |
| 6 | 332 | 10.7 | 0.00 | 0.01 | 0.01 | 0.01 |
| 7 | 327 | 10.6 | -0.10* | -0.01 | -0.03 | 0.00 |
| 8 | 320 | 10.3 | -0.18* | -0.06* | -0.03 | -0.02 |
| 9 | 289 | 9.3 | -0.27* | 0.00 | -0.07* | -0.01 |
| 10: Most deprived | 272 | 8.8 | -0.35* | 0.02 | -0.14* | -0.03 |

Note.

Sample size = 502,851 pupils in 3,098 schools.

A8 = Attainment 8.

AA8 = Adjusted Attainment 8.

P8 = Progress 8.

AP8 = Adjusted Progress 8.

* = $p < 0.05$ (statistical tests adjust for school-level clustering).